\documentclass[twocolumn,secnumarabic,amssymb, nobibnotes, aps, prd]{revtex4-1}
\RequirePackage[english]{babel}
\RequirePackage[utf8]{inputenc}
\RequirePackage{hyperref}
\hypersetup{colorlinks=true, citecolor=blue, urlcolor=blue, linkcolor=blue}

\RequirePackage[]{amsmath,amssymb}

\RequirePackage[]{graphicx}

\RequirePackage{physics}
\RequirePackage{siunitx}
\RequirePackage{cleveref}

\usepackage[dvipsnames]{xcolor}

\RequirePackage[normalem]{ulem}
\RequirePackage{orcidlink}

\begin{document}

\title{Sub-shot-noise interferometry with two mode quantum states}

\author{Quentin Marolleau\,\orcidlink{0009-0002-3587-3912}}
\author{Charlie Leprince\,\orcidlink{0009-0002-5490-6767}}
\author{Victor Gondret\,\orcidlink{0009-0005-8468-161X}}%
\author{Denis Boiron\,\orcidlink{0000-0002-2719-5931}}%
\author{Christoph I. Westbrook\,\orcidlink{0000-0002-6490-0468}}
\affiliation{Université Paris-Saclay, Institut d’Optique Graduate School, CNRS, Laboratoire Charles Fabry, 91127, Palaiseau, France}
\date{\today}%

\begin{abstract}
    We study the feasibility of sub-shot-noise interferometry with imperfect detectors, starting from twin-Fock
    states and two-mode squeezed vacuum states. We derive analytical expressions for the corresponding phase uncertainty.
    We find that one can achieve phase shift measurements below the standard quantum limit, as long as the losses are
    smaller than a given threshold, and that the measured phase is close enough to an optimal value. We provide our
    analytical formulae in a Python package, accessible online.
\end{abstract}

\maketitle

\section{Introduction}

The ability to map many physical quantities onto a phase shift makes interferometry both a crucial and generic technique
in metrology. Because of entanglement, some non-classical states can lead to improved phase resolution compared to their
classical counterparts \cite{giovannettiQuantumEnhancedMeasurementsBeating2004,giovannettiQuantumMetrology2006,
    pezzeEntanglementNonlinearDynamics2009,pezzeQuantumMetrologyNonclassical2018}.
Given an experimental resource of $N$ identical bosons, an attractive choice is to use NOON states
$\frac{1}{\sqrt{2}} \left( \ket{N,0} + \ket{0,N} \right)$. Indeed  NOON states lead to a ``Heisenberg limited" phase
uncertainty {$\Delta \phi = \mathcal{O}(N^{-1})$} \cite{bollingerOptimalFrequencyMeasurements1996a,
    dowlingCorrelatedInputportMatterwave1998,giovannettiAdvancesQuantumMetrology2011},
known to be optimal \cite{heitlerQuantumTheoryRadiation1954,ouFundamentalQuantumLimit1997}. This is a much more
advantageous scaling than the best phase sensitivity reachable with classical systems ($\Delta \phi = 1/\sqrt{N}$),
provided by a coherent state, and usually called the \emph{standard quantum limit} (SQL) or \emph{shot noise}.
Other authors have proposed the use of ``twin Fock" (TF) states $\ket{\mathrm{TF}}=\ket{N/2,N/2}$ and have shown that
they also can achieve $1/N$ scaling in phase sensitivity \cite{hollandInterferometricDetectionOptical1993,
    bouyerHeisenberglimitedSpectroscopyDegenerate1997}.

Unfortunately, NOON states are extremely fragile and behave even worse than classical states when losses are present
\cite{dunninghamInterferometryStandardQuantum2002,dornerOptimalQuantumPhase2009}. In addition, they are challenging to
prepare, and their realization with $N$ larger than a few units has not been achieved
\cite{nagataBeatingStandardQuantum2007,afekHighNOONStatesMixing2010}.

The effect of loss in quantum enhanced interferometers has been studied more generally, and states minimizing the phase
uncertainty in the presence of loss have been found \cite{demkowicz-dobrzanskiQuantumPhaseEstimation2009,
    kolodynskiPhaseEstimationPriori2010,knyshScalingLawsPrecision2011}. These states can be expressed as superpositions
of  states of the form:
\begin{equation}
    \ket{N :: m}_\pm = \frac{1}{\sqrt{2}} \left( \ket{N-m,m} \pm \ket{m,N-m} \right).
\end{equation}
Like NOON states, these states involve superpositions of strong population imbalances between the two modes (a NOON
state is in fact the special case $m=0$). This imbalance is responsible for the enhanced sensitivity, but these states
can retain their coherence despite a loss of order $m$ particles, and thus are more robust \cite{huverEntangledFockStates2008}.
In the presence of losses however, even these states can only surpass the standard quantum limit by a numerical factor,
meaning that $\Delta \phi = \mathcal{O}(N^{-1/2})$ is the best scaling possible \cite{kolodynskiPhaseEstimationPriori2010,
    escherGeneralFrameworkEstimating2011}. Here again, although the optimal states are conceptually interesting, their
experimental realization is not presently realistic.

On the other hand, the mixing on a beam splitter of the twin-Fock states mentioned above gives rise to a superposition
of $\ket{2n::2k}_\pm$ states \cite{camposQuantummechanicalLosslessBeam1989,yuspasibkoInterferenceMacroscopicBeams2014,
    laloe_quantum_2012}, and one might wonder about the robustness of such a superposition in the presence of loss. A
related state is the two-mode squeezed state (TMS) \cite{anisimovQuantumMetrologyTwoMode2010}, which is a superposition
of twin Fock states with different particle numbers. Both of these states are widely used and can be produced with a
large number of particles \cite{luoDeterministicEntanglementGeneration2017,deng2023heisenberglimited,
    andersMomentumEntanglementAtom2021,bookjansStrongQuantumSpin2011,shiskhakovMacroscopicHongOu2013,
    harderSingleModeParametricDownConversionStates2016}. These states are different from another type of experimentally
realizable state, the ``spin squeezed'' states, see fig. 5 of~\cite{pezzeQuantumMetrologyNonclassical2018}.

Other authors have studied the behaviour of twin Fock states in non-ideal interferometers \cite{tichy_double-fock_2015}.
The focus of that work was on procedures to identify different decoherence mechanisms.
Our interest here is rather to analyse how losses affect the scaling of the phase sensitivity with the number of particles.
Unlike for spin squeezed states, the relevant observable is not simply the population difference and in fact several
choices are a priori possible. We will follow other authors in using the variance of the population difference as the
interferometric observable \cite{hollandInterferometricDetectionOptical1993,bouyerHeisenberglimitedSpectroscopyDegenerate1997}.
We find that the sensitivity in this case only differs from that of the optimal state by a numerical factor and that one
can surpass the standard quantum limit if the losses are low enough.

\section{Our model}
\begin{figure}
    \includegraphics[width=\linewidth]{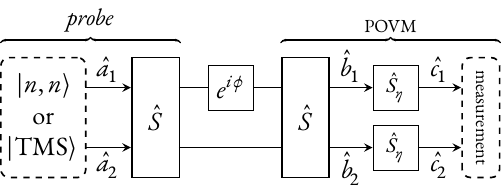}
    \caption{Generic diagram of an interferometry experiment using two-mode pure states at the input modes
        $(\hat{a}_1,\hat{a}_2)$ of the first beam splitter. After the generation of a probing state, a phase shift is
        applied, and the detection is performed (described by a positive operator valued measure). The 50:50
        beam splitter, corresponding to the unitary operator $\hat{S}$ is applied twice, and the phase difference
        between the two arms is $\phi$. The detectors have a finite quantum efficiency $\eta$, assumed to be equal,
        and is modelled with additional beam splitters $\hat{S}_\eta$ applied to the output modes of the interferometer
        $(\hat{b}_1,\hat{b}_2)$. The operators $\hat{c}_1$ and $\hat{c}_2$ represent the modes that are effectively
        detected.}
    \label{fig:schematic_MZI}
\end{figure}

We will consider the interferometer configuration represented in \cref{fig:schematic_MZI}, and for the sake of clarity we
will distinguish the \emph{input} state (before the first beam splitter) that one must prepare, from the \emph{probe}
state (after the first beam splitter) that exhibits some phase sensitivity. We assume that the losses are only caused by
the detectors, having the same quantum efficiency $\eta$ and we will consider ${\Delta \phi = 1 / \sqrt{\eta N}}$ to be
the standard quantum limit, against which we should compare our results. Our input state is either a twin Fock state or
a two-mode squeezed state. Note that without an initial beam splitter, these states produce interference patterns that
are independent of the phase \cite{hongMeasurementSubpicosecondTime1987,camposQuantummechanicalLosslessBeam1989,
    yuspasibkoInterferenceMacroscopicBeams2014}.

Two-mode squeezed states are spontaneously generated from vacuum with a quadratic interaction Hamiltonian. By denoting
${\xi = r e^{i \theta}}$ the squeezing parameter, whose norm $r$ is proportional to the interaction time, such a state
reads
\begin{equation}
    \ket{\mathrm{TMS}} = \frac{1}{\cosh(r)} \sum\limits_{n=0}^\infty e^{i n \theta} \tanh^n(r) \ket{n,n}
\end{equation}
in the Fock basis relative to the modes $\hat{a}_1$ and $\hat{a}_2$ (see \cref{fig:schematic_MZI}).

The action of the interferometer on the input state is described by the unitary operator $\hat{U}$:
\begin{equation}
    \hat{U} = \hat{S} \begin{pmatrix}
        e^{i\phi} & 0 \\ 0 & 1
    \end{pmatrix}     \hat{S} = e^{i \frac{\phi}{2}}
    \begin{bmatrix}
        i \sin(\frac{\phi}{2}) & \cos(\frac{\phi}{2})    \\
        -\cos(\frac{\phi}{2})  & -i \sin(\frac{\phi}{2})
    \end{bmatrix}
\end{equation}
The losses are modelled by additional beam splitters $\hat{S}_\eta$ placed at the output ports:
\begin{equation}
    \hat{S}_\eta =
    \begin{bmatrix}
        \sqrt{\eta}     & \sqrt{1-\eta} \\
        - \sqrt{1-\eta} & \sqrt{\eta}
    \end{bmatrix}
\end{equation}
With the input and output annihilation operators defined in \cref{fig:schematic_MZI}, we introduce the additional
notations for the number operators:
\begin{equation}
    \hat{N}_{\alpha_i} = \hat{\alpha}^\dagger_i \hat{\alpha}_i
    \quad / \quad \alpha \in \{a,b,c \} \ , \ i \in\{ 1,2 \}
\end{equation}
and the detected particle number difference at the output, with and without losses:
\begin{equation}
    \left\{
    \begin{aligned}
        \hat{D}_\eta & = \frac{1}{2} \left( \hat{N}_{c_2} - \hat{N}_{c_1} \right)                      \\
        \hat{D}      & = \frac{1}{2} \left( \hat{N}_{b_2} - \hat{N}_{b_1} \right) = \hat{D}_{\eta = 1}
    \end{aligned}
    \right.
\end{equation}

We also denote ${N=\ev{\hat{N}_{a_1}+\hat{N}_{a_2}}=\ev{\hat{N}_{b_1}+\hat{N}_{b_2}}}$ the average number of particles
in the initial state. In the case of a twin Fock state $\ket{n,n}$, we simply have $N=2n$, whereas for a two-mode
squeezed state $N = 2 \sinh^2(r)$, i.e. twice the average number of particles per mode. The mean number of detected
particles therefore is $\eta N$.

The operator $\hat{U}$ provides the expansion of $\hat{D}$ in terms of the input modes:
\begin{equation}
    \hat{D} =\frac 12 \left[\cos(\phi) \, \left( \hat{N}_{a_1} - \hat{N}_{a_2} \right) +i \sin(\phi) \, \left( \hat{a}^\dagger_1 \hat{a}_2  - \hat{a}_2^\dagger \hat{a}_1\right) \right]
\end{equation}
Therefore, whatever the phase $\phi$ and the quantum efficiency $\eta$, a twin Fock state placed at the input of the
interferometer yields a vanishing expectation value for $\hat{D}$. Due to linearity, the same is true for two-mode
squeezed states. This means that $\hat{D}$ itself is not a suitable observable to recover information about the phase
$\phi$ with those states. However one can study  $\hat{D}^2$ which characterizes the width of these distributions.

Indeed, one can derive \cite{supplemental}:
\begin{equation}
    \left\{
    \begin{aligned}
        \ev{\hat{D}_\eta^2}_{\mathrm{tf}}  & = \eta^2 \frac N4 \left(1+\frac N2\right)\sin^2(\phi) + \eta (1-\eta) \frac N4 \\
        \ev{\hat{D}_\eta^2}_{\mathrm{tms}} & = \eta^2 \frac N2 \left(1+\frac N2\right)\sin^2(\phi) + \eta (1-\eta) \frac N4
    \end{aligned}
    \right.
\end{equation}
making explicit the phase dependence. The $\eta (1-\eta) \frac N4$ offset is the Poissonian noise due to the fact that
the losses at each detector are uncorrelated.

\section{Results}
The phase uncertainty can be computed analytically using
\begin{equation}
    \Delta \phi =\frac{\sqrt{\mathrm{Var}\left[\hat{D}_\eta^2\right]}}{\left|\dfrac{\partial}{\partial \phi}\left[ \ev{\hat{D}_\eta^2} \right] \right|}.
    \label{eq:phase_uncertainty}
\end{equation}
If the detectors are lossless ($\eta =1$), the phase uncertainties are given by:
\begin{equation}
    \begin{cases}
        \Delta \phi_{\mathrm{tf}} = \dfrac{1}{\cos (\phi ) \sqrt{N (N+2)}} \sqrt{2 + \left(-3 + \frac N4 + \frac{N^2}{8}\right) \sin ^2(\phi )} \\
        \Delta \phi_{\mathrm{tms}} = \dfrac{1}{\cos (\phi ) \sqrt{N (N+2)}} \sqrt{1 + 2 N (N+2) \sin ^2(\phi )}
    \end{cases}
    \label{eqlossless}
\end{equation}
In the neighbourhood of ${\phi = 0}$, we find Heisenberg limited scaling ${\Delta \phi = \mathcal{O}(N^{-1})}$. We also
note that the TMS state outperforms the TF state by a factor of $\sqrt{2}$.

\begin{figure}
    \includegraphics[width=\linewidth]{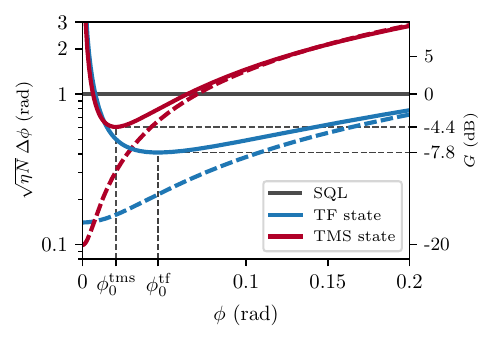}
    \caption{Ratio between the phase uncertainty $\Delta \phi$ and the SQL, using respectively twin-Fock state (in blue)
        and two-mode squeezed
        state (in red) as input states of the interferometer. The dashed lines correspond to the situation
        where the quantum efficiency of the detectors is assumed to be perfect ($\eta=1$), whereas the plain lines refer
        to detectors with finite quantum efficiency (here $\eta=0.95$). Both types of states have an average population
        of 100 particles (50 per mode). We also give the gain in decibel defined as $G = 20 \log \left(\sqrt{\eta N} \,
            \Delta\phi\right)$.}
    \label{fig:phase_uncertainty}
\end{figure}
\begin{figure}
    \includegraphics[width=\linewidth]{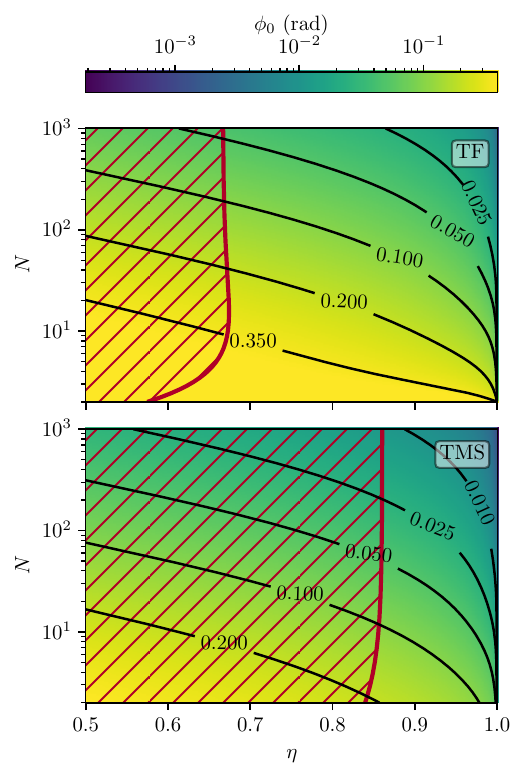}
    \caption{Optimal phase $\phi_0$ that maximises the phase resolution during a measurement, plotted as a function
        of the number of particles in the interferometer $N$ and the quantum efficiency of the detectors $\eta$.
        The red hatches exhibit the subdomain of the $(\eta, N)$ plane where no measurement below the SQL can be performed.
        Isolines of $\phi_0$ are plotted in black. The top graph represents $\phi_0$ for the TF state, while the bottom
        graph refers to the TMS state.}
    \label{fig:optimal_phi}
\end{figure}

When we include losses, the analogous expressions become rather long and we leave them to the supplementary materials
\cite{supplemental}. In the example in \cref{fig:phase_uncertainty}, we show that the phase uncertainty $\Delta \phi$
can be smaller than the standard quantum limit. In addition, the phase uncertainty has a minimum at a non-zero phase
$\phi_0$ which depends on the detection efficiency, number of particles and the input state (see \cref{fig:optimal_phi}).
The optimal phase is shifted because, in presence of losses, $\mathrm{Var}\left[\hat{D}_\eta^2\right]$ does not vanish
at $\phi=0$, leading to a divergence ${\Delta \phi \underset{\phi = 0}{=} \mathcal{O}(\phi^{-1})}$. Also, unlike the
lossless case, in this situation the twin-Fock state outperforms the two-mode squeezed state even at the optimal phase
$\phi_0$ \cite{supplemental}.

This type of profile has been observed experimentally \cite{luckeTwinMatterWaves2011}. From the study of the variations
of $\Delta \phi$ as a function of $\phi$, one can compute the optimal phase $\phi_0$ around which an experiment should
operate to perform precision measurements. This means that during an experiment, one must be able to tune a phase offset,
for instance in optics by adding a tiltable glass plate.

In \cref{fig:optimal_phi} we show color maps of the values of $\phi_0$ as a function of the number of particles $N$ and
the quantum efficiency $\eta$. Regions where sub-shot-noise measurements are possible correspond to non-hashed regions.
It appears in these maps that this question is mostly related to the quantum efficiency of the detectors: depending on
whether one is dealing with twin Fock or two-mode squeezed states, a threshold of $\eta \approx 0.7$ or respectively
$\eta \approx 0.9$ must be achieved to surpass the standard quantum limit.

\begin{figure}
    \includegraphics[width=\linewidth]{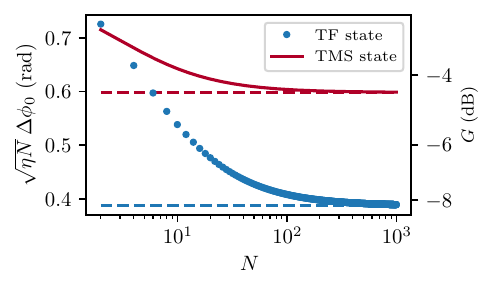}
    \caption{Asymptotic behavior of the ratio between the phase uncertainty $\Delta \phi_0$ (i.e. $\Delta \phi$ estimated
        at the optimal phase $\phi_0$) and the SQL, as a function of the number of particles, for both TF and TMS states.
        In the case of TF states, the number of particles $N$ is restricted to even integers. The quantum efficiency is
        set to ${\eta = 0.95}$. The point at $N=2$ showing a TMS sensitivity better than TF is an illustration of the
        points made below \cref{eqlossless} that in the lossless case the TMS performs slightly better but that this
        advantage disappears with decreasing $\eta$ and increasing $N$ (see \cite{supplemental}). Unlike the lossless
        case which provides an $N^{-1}$ scaling, the ratios converge towards a finite limit (dashed line), meaning that
        the SQL is surpassed only by a constant factor. We also give the gain in decibel defined as $G = 20 \log
            \left(\sqrt{\eta N} \, \Delta\phi_0\right)$.
    }
    \label{fig:delta_phi_phi0}
\end{figure}

We have computed $\Delta\phi_0$ \cite{supplemental}, the phase uncertainty when the measurement is performed at the
optimal phase $\phi_0$. When $N$ is small, $\Delta\phi_0$ varies similarly to a power law ${\Delta\phi_0 \approx 1 / N^\alpha}$
with ${0.5 < \alpha < 1}$, depending on the value of $\eta$ (an example is plotted in \cref{fig:delta_phi_phi0}).
Experimentally, in this region one obtains significant gains with respect to the standard quantum limit by increasing
the number of particles. In the asymptotic region, where $N$ goes to infinity, we recover the ${\Delta\phi_0 = \mathcal{O}(N^{-1/2})}$ scaling \cite{kolodynskiPhaseEstimationPriori2010,escherGeneralFrameworkEstimating2011}.

\begin{figure}
    \includegraphics[width=\linewidth]{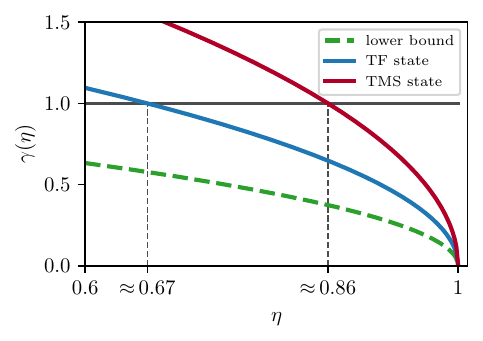}
    \caption{Ratio between the phase uncertainty $\Delta \phi_0$ and the SQL, in the asymptotic regime, as a function of
        the quantum efficiency (cf. \cref{eq:def_gamma}). We find that both TF and TMS states give a profile which is
        proportional to the one obtained with an optimized input state (green dashed line). We also see in this graph
        the minimal values of $\eta$ leading to sub-shot-noise measurements (these values correspond to the limit
        ${N \rightarrow \infty}$ of the red lines in \cref{fig:optimal_phi}).}
    \label{fig:res_limit}
\end{figure}

We also compute
\begin{equation}
    \gamma(\eta) = \lim\limits_{N\rightarrow \infty}\sqrt{\eta N} \, \Delta\phi_0
    \label{eq:def_gamma}
\end{equation}
which is the ratio between $\Delta \phi_0$ and the standard quantum limit, in the asymptotic limit. This quantity tells
what value of $\eta$ must be reached to go below the SQL. It has been proven \cite{kolodynskiPhaseEstimationPriori2010}
that
\begin{equation}
    \sqrt{\eta N} \, \Delta\phi \geq \sqrt{1-\eta}
    \label{eq:lower_bound_gamma}
\end{equation}
but this bound is tight only when using optimal input states, as well as an observable which is not explicitly known.
In our case, the function $\gamma$ is actually a simple dilation of the lower bound (\cref{eq:lower_bound_gamma}) (see
\cref{fig:res_limit}):
\begin{equation}
    \left\{
    \begin{aligned}
        \gamma^{\mathrm{tf}}(\eta)  & = \sqrt{3} \, \sqrt{1-\eta}                                                                                                                                           \\
        \gamma^{\mathrm{tms}}(\eta) & = \vphantom{\left(\frac{2}{5}\right)^{1/4}} \smash{\underbrace{\left(\frac{2}{5}\right)^{1/4} \sqrt{5 + 2 \sqrt{10}}}_{\displaystyle \approx 2.676}} \, \sqrt{1-\eta}
    \end{aligned}
    \right.
    \vspace*{3ex}
    \label{eq:lowerbondTFTMS}
\end{equation}
Our simple measurement protocol is therefore similar to an optimal situation where ideal states are used.

\newpage
\section{Conclusion}

We have shown that two classes of experimentally accessible states, twin-Fock and two-mode squeezed states can behave in
a way reminiscent of ideal ones with respect to their phase sensitivity in interferometers. In the absence of any loss,
their phase sensitivity exhibits $\mathcal{O}(N^{-1})$ scaling, as do NOON states. On the other hand, in the presence of
loss, they are more robust than NOON states. Although the sensitivity scales only as $\mathcal{O}(N^{-1/2})$ they can
still surpass the standard quantum limit if the losses are kept small enough. In this sense they resemble other ideal
states which have been shown to be optimal in the presence of loss, only differing from the optimal states by a
numerical factor, see \cref{eq:lowerbondTFTMS}. We show in \cref{fig:delta_phi_phi0} that using a twin-Fock state and a
95\% quantum efficiency results in a 8 dB improvement compared to the standard quantum limit which is not very far from
the theoretical bound of 13 dB given by \cref{eq:lowerbondTFTMS}. For a two-mode squeezed state the gain is only 4.4 dB.

We emphasize that minimizing the loss is critical. Fig.~\ref{fig:res_limit} shows that both types of state require a
minimum quantum efficiency to achieve a gain. Fig.~\ref{fig:optimal_phi} shows that this minimum is roughly independent
of the number of particles on that state. Despite this drawback, we expect that such improvements could be useful in
some interferometers: for example when increasing the number of particles to reduce the shot noise is not practical.
Whether twin Fock or two-mode squeezed states constitute a real advantage compared to spin squeezing will require more
work in the future using comparisons in realistic experimental situations \cite{pezzeQuantumMetrologyNonclassical2018}.
The fact that these relatively accessible states are not far from the optimized ones is an encouraging sign.

Our analytical formulae are provided in the supplementary materials and are implemented in a Python package, accessible
online at \url{https://github.com/quentinmarolleau/qsipy}.

\begin{acknowledgments}
    The research leading to these results has received funding from QuantERA Grant No. ANR-22-QUA2-0008-01 (MENTA) and
    ANR Grant No. 20-CE-47-0001-01 (COSQUA), the LabEx PALM, Région Ile-de-France in the framework of the DIM SIRTEQ
    program and Quantum Paris-Saclay.
\end{acknowledgments}
\clearpage

\bibliography{bibliography}

\end{document}


\title{Supplemental material for \\ \textit{Interferometric phase estimation with two mode quantum states}}

\author{Quentin Marolleau}
\author{Charlie Leprince}
\author{Victor Gondret}%
\author{Denis Boiron}%
\author{Christoph I. Westbrook}
\affiliation{Université Paris-Saclay, Institut d’Optique Graduate School, CNRS, Laboratoire Charles Fabry, 91127, Palaiseau, France}
\date{\today}%

\maketitle

\tableofcontents
\pagebreak

\section{Parametrization of the problem}

\subsection{Operator definitions}

\begin{figure}[h]
	\includegraphics{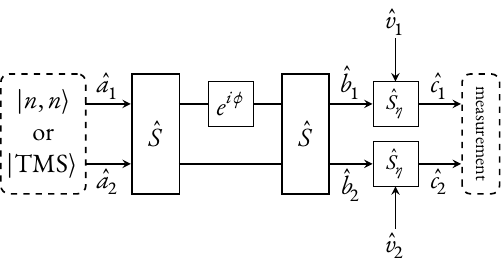}
	\caption{Scheme of the interferometer that we study. $\hat{a}_1$ and $\hat{a}_2$ are the input modes, 50:50 beam splitters, corresponding to the unitary operator $\hat{S}$ are applied twice, and the phase difference between the two arms is $\phi$. The detectors placed at the output ports have a finite quantum efficiency $\eta$, assumed to be the same, and that is modelled with additional beam splitters
		$\hat{S}_\eta$ applied to the output modes of the interferometer $(\hat{b}_1,\hat{b}_2)$. The operators
		$\hat{c}_1$ and $\hat{c}_2$ represent the modes that are effectively detected. The modes $\hat{v}_1$ and
		$\hat{v}_2$ represent the vacuum channels.}
	\label{fig:schematic_MZI}
\end{figure}

Following the notation of \cref{fig:schematic_MZI}, we denote $N$ the average value of the total number of atoms in the
interferometer:
\begin{equation}
	N \triangleq \ev{\hat{N}_{a_1} + \hat{N}_{a_2}} = \ev{\hat{N}_{b_1} + \hat{N}_{b_2}}
\end{equation}
We have
\begin{equation}
	\hat{S} = \frac{1}{\sqrt{2}} \begin{pmatrix}
		1  & 1 \\
		-1 & 1
	\end{pmatrix}
\end{equation}
corresponding to the special case of a beam splitter that does not apply any phase shift.
\begin{equation}
	\hat{\Phi} = \begin{pmatrix}
		e^{i \phi} & 0 \\
		0          & 1
	\end{pmatrix}
\end{equation}
\begin{equation}
	\hat{U} = \hat{S} \hat{\Phi} \hat{S} = e^{i \frac{\phi}{2}} \begin{pmatrix}
		i \sin(\frac{\phi}{2}) & \cos(\frac{\phi}{2})    \\
		-\cos(\frac{\phi}{2})  & -i \sin(\frac{\phi}{2})
	\end{pmatrix}
\end{equation}

\begin{equation}
	\binom{\hat{b}_1}{\hat{b}_2} = \hat{U} \binom{\hat{a}_1}{\hat{a}_2}
\end{equation}
We also have the beam splitters modelling the losses:
\begin{equation}
	\hat{S}_\eta =
	\begin{pmatrix}
		\sqrt{\eta}     & \sqrt{1-\eta} \\
		- \sqrt{1-\eta} & \sqrt{\eta}
	\end{pmatrix}
\end{equation}
such that
\begin{equation}
	\hat{c}_i = \sqrt{\eta} \, \hat{b}_i + \sqrt{1-\eta} \, \hat{v}_i \quad / \quad i \in \{ 1,2 \}
\end{equation}
We finally introduce notations for the number operators:
\begin{equation}
	\hat{N}_{\alpha_i} = \hat{\alpha}^\dagger_i \hat{\alpha}_i \quad / \quad  \alpha \in \{a,b,c \} \ , \ i \in\{ 1,2 \}
\end{equation}
input spin operators:
\begin{equation}
	\begin{cases}
		\hat{J}_x =\frac 12 \left( \hat{a}^\dagger_1 \hat{a}_2  + \hat{a}_2^\dagger \hat{a}_1\right)    \\
		\hat{J}_y =\frac 1{2i} \left( \hat{a}^\dagger_1 \hat{a}_2  - \hat{a}_2^\dagger \hat{a}_1\right) \\
		\hat{J}_z =\frac 12 \left( \hat{N}_{a_1} - \hat{N}_{a_2} \right)
	\end{cases}
\end{equation}
and the observables of interest:
\begin{equation}
	\left\{
	\begin{aligned}
		\hat{D}_\eta & = \frac{1}{2} \left( \hat{N}_{c_2} - \hat{N}_{c_1} \right)                      \\
		\hat{D}      & = \frac{1}{2} \left( \hat{N}_{b_2} - \hat{N}_{b_1} \right) = \hat{D}_{\eta = 1}
	\end{aligned}
	\right.
\end{equation}

\subsection{Two-mode squeezed vacuum state and preliminary results}

We recall the definition of a two-mode squeezed vacuum (TMS) state, with average total population $N$:
\begin{equation}
	\ket{\textrm{TMS}} \triangleq \sqrt{ \frac{2}{2+N}} \sum\limits_{n=0}^{\infty} \left( \frac{N}{2+N} \right)^{\frac n2} \ket{n,n}
\end{equation}
in order to keep compact notations during the calculations, we will often use the \emph{thermal weight}:
\begin{equation}
	P_{th}(n) = \frac{2}{2+N}\left( \frac{N}{2+N} \right)^n
\end{equation}
corresponding to the probability to measure $n$ particles in a given mode of the TMS state.

\paragraph*{} We also highlight the fact that
\begin{equation}
	\hat{J}_z \ket{n,n} = \hat{J}_z \ket{\mathrm{TMS}} = 0
	\label{eq:jz_gives_zero}
\end{equation}
and finally:
\begin{equation}
	\ev{\hat{J}_x} = \ev{\hat{J}_y} = 0
	\label{eq:ev_jx_jy_is_zero}
\end{equation}
for both twin Fock and two-mode squeezed vacuum states.

\section{Expansion of $\hat{D}$, $\hat{D}^2$, $\hat{D}_\eta$ and $\hat{D}^2_\eta$}

\begin{equation}
	\hat{D} = \cos (\phi) \hat{J}_z - \sin (\phi) \hat{J}_y
\end{equation}

\begin{equation}
	\hat{D}^2 = \cos^2 (\phi) \hat{J}_z^2 + \sin^2 (\phi) \hat{J}_y^2 - 2 \sin (\phi) \cos (\phi) \hat{J}_y \hat{J}_z
	+ i \sin (\phi) \cos (\phi) \hat{J}_x
	\label{eq:expansion_of_D2}
\end{equation}
Since there is no particle in the vacuum channels for the input state, we always have $\hat{v}_i \ket{\psi}_{\textrm{input}} = 0$.
For the sake of simplicity, we reduce the writing of $\hat{D}_\eta$ and $\hat{D}^2_\eta$ to the terms giving  a non zero
contribution. This means that (for either $\hat{D}_\eta$ and $\hat{D}_\eta^2$) we drop all the terms containing $\hat{v}_{i \in \{ 1,2 \}}$ annihilation operators on their
rightmost side:
\begin{equation}
	\hat{D}_\eta = \eta \hat{D} + \frac 12 \sqrt{\eta (1-\eta)} \left( \hat{v}^\dagger_2 \, \hat{b}_2 - \hat{v}^\dagger_1 \, \hat{b}_1 \right)
\end{equation}

\begin{equation}
	\begin{split}
		\hat{D}_\eta^2 = \eta^2 \hat{D}^2 + \frac {\eta(1-\eta)}{4} \left[ (\hat{v}^\dagger_1)^2 \, \hat{b}_1^2
			+ (\hat{v}^\dagger_2)^2 \, \hat{b}_2^2 + \hat{N}_{b_1}+\hat{N}_{b_2} - 2 \, \hat{b}_1 \hat{b}_2 \, \hat{v}^\dagger_1 \hat{v}^\dagger_2 \right] \\
		+ \frac{\eta \sqrt{\eta (1-\eta)}}{2} \left[ 2 \hat{D} \, \hat{b}_2 \hat{v}_2^\dagger + \frac 12 \hat{b}_2 \hat{v}_2^\dagger
			- 2 \hat{D} \, \hat{b}_1 \hat{v}_1^\dagger + \frac 12 \hat{b}_1 \hat{v}_1^\dagger \right] \\
		+ \frac{(1-\eta) \sqrt{\eta (1-\eta)}}{4} \left[ \left( 2 \hat{N}_{v_2} - \mathbbm{1} \right) \hat{b}_2 \hat{v}^\dagger_2
			+ \left( 2 \hat{N}_{v_1} - \mathbbm{1} \right) \hat{b}_1 \hat{v}^\dagger_1 \right]
	\end{split}
	\label{eq:D_eta_2}
\end{equation}

\section{Expectation values of $\hat{D}^2$ and $\hat{D}^2_\eta$}

\subsection{Lossless case}

\subsubsection*{With twin Fock states}

\begin{equation}
	\hat{J}_y^2 = \frac 14 \left[ \hat{N}_{a_1} \left( \mathbbm{1} + \hat{N}_{a_2} \right)
		+ \hat{N}_{a_2} \left( \mathbbm{1} + \hat{N}_{a_1} \right) - ( \hat{a}^\dagger_1 )^2 \, \hat{a}_2^2
		- ( \hat{a}^\dagger_2 )^2 \, \hat{a}_1^2 \right]
\end{equation}
thus with \eqref{eq:jz_gives_zero} \eqref{eq:ev_jx_jy_is_zero} and \eqref{eq:expansion_of_D2},
\begin{equation}
	\ev{\hat{D}^2}_{\mathrm{tf}} = \ev{\hat{J}_y^2}_{\mathrm{tf}} \sin^2 (\phi) = \frac{N}{4} \left( 1 + \frac{N}{2} \right) \sin^2 (\phi)
	\label{eq:d2_perfect_tf}
\end{equation}

\subsubsection*{With two-mode squeezed vacuum states}

We can check that
\begin{equation}
	m \neq n \Rightarrow \mel**{n,n}{\hat{D}^2}{m,m} = 0
\end{equation}
therefore assuring the simple relation:
\begin{equation}
	\ev{\hat{D}^2}_{\mathrm{tms}} = \sum\limits_{n=0}^\infty P_{th}(n) \ev{\hat{D}^2}_{\mathrm{tf}}
\end{equation}
leading to
\begin{equation}
	\ev{\hat{D}^2}_{\mathrm{tms}} = \frac N2 \left( 1 + \frac N2 \right) \sin^2 (\phi) = 2 \ev{\hat{D}^2}_{\mathrm{tf}}
	\label{eq:d2_perfect_tms}
\end{equation}

\subsection{Lossy case (i.e. eq. (8) in the main paper)}

\subsubsection*{With twin Fock states}

\vspace*{1ex}
Whatever the output state of the interferometer, we actually always have:
\begin{equation}
	\mathrm{Var}\left[ \hat{N}_{c_2} - \hat{N}_{c_1} \right] =
	\eta^2 \, \mathrm{Var}\left[ \hat{N}_{b_2} - \hat{N}_{b_1} \right]
	+ \eta (1-\eta) \, \ev{\hat{N}_{b_2} + \hat{N}_{b_1}}
\end{equation}
which in our case means:
\begin{equation}
	\ev{\hat{D}_\eta^2}_{\mathrm{tf}} = \eta^2 \ev{\hat{D}^2}_{\mathrm{tf}} + \frac{\eta (1- \eta)}{4} N
	\label{eq:d2_eta_tf}
\end{equation}
and therefore
\begin{equation}
	\ev{\hat{D}_\eta^2}_{\mathrm{tf}} = \eta^2 \frac N4 \left( 1 + \frac N2 \right) \sin^2 (\phi) + \frac{\eta (1- \eta)}{4} N
\end{equation}

\subsubsection*{With two-mode squeezed vacuum states}

Again we can check on \cref{eq:D_eta_2} that
\begin{equation}
	m \neq n \Rightarrow \mel**{n,n}{\hat{D}_\eta^2}{m,m} = 0
\end{equation}
thus we still have
\begin{equation}
	\ev{\hat{D}_\eta^2}_{\mathrm{tms}} = \sum\limits_{n=0}^\infty P_{th}(n) \ev{\hat{D}_\eta^2}_{\mathrm{tf}}
\end{equation}
and finally:
\begin{equation}
	\ev{\hat{D}_\eta^2}_{\mathrm{tms}} = \eta^2 \frac N2 \left( 1 + \frac N2 \right) \sin^2 (\phi) + \frac{\eta (1- \eta)}{4} N
	\label{eq:d2_eta_tms}
\end{equation}

\section{Expectation values of $\hat{D}^4$ and $\hat{D}^4_\eta$}

\subsection{Lossless case}

\subsubsection*{With twin Fock states}

We compute  $\ev{\hat{D}^4}_{\mathrm{tf}} = \norm{\hat{D}^2 \ket{\frac N2, \frac N2}}^2$. The only
non-vanishing term of $\hat{D}^2 \ket{n,n}$ are:
\begin{equation}
	\begin{cases}
		\displaystyle \hat{J}_y^2 \ket{n,n} & \displaystyle = \frac 14 \left( 2n \left( 1 + n\right) \ket{n,n}

		- \sqrt{\left( n - 1 \right) n \left( n + 1 \right) \left( n + 2 \right)} \big[ \ket{n+2,n-2} + \ket{n-2,n+2} \big]\right) \\
		\\
		\displaystyle \hat{J}_x \ket{n,n}   & = \displaystyle \frac 12 \sqrt{n(n+1)} \big( \ket{n+1,n-1} + \ket{n-1,n+1} \big)
	\end{cases}
	\label{eq:Jy2_Jx_system}
\end{equation}
all these vectors are mutually orthogonal, then:
\begin{equation}
	\norm{\hat{D}^2 \ket{\frac N2, \frac N2}}^2 = \frac{N}{4} \left( 1 + \frac N2 \right) \sin^2 (\phi) \left[ 1+ \frac{3}{2}
		\left( -1 + \frac N4 + \frac{N^2}{8} \right) \sin^2 (\phi) \right]
	\label{eq:d4_perfect_tf}
\end{equation}

\subsubsection*{With two-mode squeezed vacuum states}

Looking at \cref{eq:Jy2_Jx_system}, we can convince ourselves that the decomposition in the two-mode Fock basis of
$\hat{J}_y^2 \ket{n,n}$ and $\hat{J}_x \ket{n,n}$ with $n \in \mathbb{N}$ generate vectors that are all mutually orthogonal.

Therefore
\begin{equation}
	\norm{\hat{D}^2 \ket{\mathrm{TMS}}}^2 = \sum\limits_{n=0}^\infty P_{th}(n) \norm{\hat{D}^2 \ket{n, n}}^2
\end{equation}
and thus,
\begin{equation}
	\norm{\hat{D}^2 \ket{\mathrm{TMS}}}^2 = \frac{N}{2} \left( 1 + \frac N2 \right) \sin^2 (\phi) \left[ 1+ \frac{9N}{2}
		\left( 1 + \frac N2 \right) \sin^2 (\phi) \right]
	\label{eq:d4_perfect_tms}
\end{equation}

\subsection{Lossy case}

We follow the same procedure as before, but here the number of non-vanishing terms is much larger. We will only write
the final results.

\subsubsection*{With twin Fock states}

With
\begin{equation}
	\begin{cases}
		P^{\mathrm{tf}}_0(N,\eta) & = 64 - 320 \, \eta + 256 \, \eta N + 384 \, \eta^2 - 384 \, \eta^2 N + 96 \, \eta^2 N^2 - 144 \, \eta^3 + 156 \, \eta^3 N - 60 \, \eta^3 N^2 + 9 \, \eta^3 N^3 \\
		P^{\mathrm{tf}}_1(N,\eta) & = - 4 \, (N+2) \, \eta \left( 16 + 24 \, (N-2) \, \eta + 3 \left( 8 - 6N + N^2 \right) \eta^2 \right)                                                          \\
		P^{\mathrm{tf}}_2(N,\eta) & = 3 \left( -16 - 4N + 4N^2 + N^3 \right) \eta^3
	\end{cases}
\end{equation}
we have:
\begin{equation}
	\norm{\hat{D}_\eta^2 \ket{\frac N2, \frac N2}}^2 = \frac{\eta N}{1024}
	\Big[ P^{\mathrm{tf}}_0(N,\eta) + P^{\mathrm{tf}}_1(N,\eta) \cos (2\phi) + P^{\mathrm{tf}}_2(N,\eta) \cos(4\phi) \Big]
	\label{eq:d4_eta_tf}
\end{equation}

\subsubsection*{With two-mode squeezed vacuum states}

With
\begin{equation}
	\begin{cases}
		P^{\mathrm{tms}}_0(N,\eta) & = 8 + 24 \, \eta + 64 \, \eta N + 48 \, \eta^2 N + 72 \, \eta^2 N^2 + 12 \, \eta^3 N + 36 \, \eta^3 N^2 + 27 \, \eta^3 N^3 \\
		P^{\mathrm{tms}}_1(N,\eta) & = - 4 \, (N+2) \, \eta \left( 4 + 18 \, \eta N + 9 \, \eta^2 N^2 \right)                                                   \\
		P^{\mathrm{tms}}_2(N,\eta) & = 9N \left( N+2 \right)^2 \eta^3
	\end{cases}
\end{equation}
we have:
\begin{equation}
	\norm{\hat{D}_\eta^2 \ket{\mathrm{TMS}}}^2 = \frac{\eta N}{128}
	\Big[ P^{\mathrm{tms}}_0(N,\eta) + P^{\mathrm{tms}}_1(N,\eta) \cos (2\phi) + P^{\mathrm{tms}}_2(N,\eta) \cos(4\phi) \Big]
	\label{eq:d4_eta_tms}
\end{equation}

\section{Phase uncertainty $\Delta \phi$}

Phase uncertainties are computing using
\begin{equation}
	\Delta \phi =\frac{\sqrt{\mathrm{Var}\left[\hat{D}_\eta^2\right]}}{\left|\dfrac{\partial}{\partial \phi}\left[ \ev{\hat{D}_\eta^2} \right] \right|}.
	\label{eq:def_delta_phi}
\end{equation}

\subsection{Lossless case (i.e. eq. (10) in the main paper)}

\subsubsection*{With twin Fock states}

Injecting \eqref{eq:d2_perfect_tf} and \eqref{eq:d4_perfect_tf} into \cref{eq:def_delta_phi}, we get:
\begin{equation}
	\Delta \phi_{\mathrm{tf}} = \dfrac{1}{\cos (\phi ) \sqrt{N (N+2)}} \sqrt{2 + \left(-3 + \frac N4 + \frac{N^2}{8}\right) \sin ^2(\phi )}
\end{equation}

\subsubsection*{With two-mode squeezed vacuum states}

Injecting \eqref{eq:d2_perfect_tms} and \eqref{eq:d4_perfect_tms} into \cref{eq:def_delta_phi}, we get:
\begin{equation}
	\Delta \phi_{\mathrm{tms}} = \dfrac{1}{\cos (\phi ) \sqrt{N (N+2)}} \sqrt{1 + 2 N (N+2) \sin ^2(\phi )}
\end{equation}

\subsection{Lossy case}

\subsubsection*{With twin Fock states}


Injecting \eqref{eq:d2_eta_tf} and \eqref{eq:d4_eta_tf} into \cref{eq:def_delta_phi}, we get:

\begin{equation}
	\begin{cases}
		Q^{\mathrm{tf}}_0(N,\eta) & = -144 \eta ^3+384 \eta ^2-320 \eta +3 \eta ^3 N^3-52 \eta ^3 N^2+64 \eta ^2 N^2+132 \eta ^3 N-320 \eta ^2 N+192 \eta  N+64 \\
		Q^{\mathrm{tf}}_1(N,\eta) & = -4 \eta \, (N+2) \left(\eta ^2 \left(N^2-14 N+24\right)+16 \eta \, (N-3)+16\right)                                        \\
		Q^{\mathrm{tf}}_2(N,\eta) & = \eta ^3 \left(N^3+4 N^2-20 N-48\right)
	\end{cases}
\end{equation}
\begin{equation}
	\begin{split}
		\Delta \phi = \frac{1}{4N(N+2) \, \eta^2 \, \left|\sin(2 \phi)\right|} \sqrt{\eta \, N \left[Q^{\mathrm{tf}}_0(N,\eta) + Q^{\mathrm{tf}}_1(N,\eta) \cos(2\phi) + Q^{\mathrm{tf}}_2(N,\eta) \cos(4\phi)\right]}
	\end{split}
\end{equation}

\subsubsection*{With two-mode squeezed vacuum states}

Injecting \eqref{eq:d2_eta_tms} and \eqref{eq:d4_eta_tms} into \cref{eq:def_delta_phi}, we get:
\begin{equation}
	\begin{cases}
		Q^{\mathrm{tms}}_0(N,\eta) & = 3 \eta +3 \eta ^3 N^3+4 \eta ^3 N^2+8 \eta ^2 N^2+\eta ^3 N+6 \eta ^2 N+7 \eta  N+1 \\
		Q^{\mathrm{tms}}_1(N,\eta) & = -2 \eta \, (N+2) \left(2 \eta ^2 N^2+4 \eta  N+1\right)                             \\
		Q^{\mathrm{tms}}_2(N,\eta) & = \eta ^3 N (N+2)^2
	\end{cases}
\end{equation}
\begin{equation}
	\begin{split}
		\Delta \phi = \frac{1}{N(N+2) \, \eta^2 \, \left|\sin(2 \phi)\right|} \sqrt{\eta \, N \left[Q^{\mathrm{tms}}_0(N,\eta) + Q^{\mathrm{tms}}_1(N,\eta) \cos(2\phi) + Q^{\mathrm{tms}}_2(N,\eta) \cos(4\phi)\right]}
	\end{split}
\end{equation}

\section{Optimal phase $\phi_0$}

When considering non-unit quantum efficiency, the phase uncertainty exhibits a minimum in $\phi_0 > 0$. In that case,
the study of the derivative of $\Delta \phi$ as a function of $\phi$ gives the analytic expression of $\phi_0$.

\subsubsection*{With twin Fock states}

\begin{equation}
	\begin{split}
		\phi_0 = \arccsc \Bigg[ \bigg(\eta ^3 N (N^2-12 N+12)-2 \sqrt{2} \Big(\\
			&(\eta -1) (-16 \eta ^2 (5 N^2-19 N+12) \\
			&+2 \eta ^5 N (N^3-15 N^2+48 N-36) \\
			&-2 \eta ^4 N (N^3-31 N^2+144 N-180) \\
			&+\eta ^3 (-33 N^3+268 N^2-540 N+144) \\
			&+\eta  (72-48 N)-8) \\
			&\Big)^{1/2}\\
			+16 \eta ^2 (N-3) N+8 \eta  (4 N-3)+8\bigg)^{1/2} \Bigg]
	\end{split}
\end{equation}

\section{Comparison of TF and TMS state in the presence of losses}

In the main paper, we noticed that when considering unit quantum efficiency, the TMS
outperforms the TF by a factor of $\sqrt{2}$, in the neighbourhood of the optimal phase $\phi_0 =0$
(see eq. (10) of the main paper).

However, it is questionable to conclude that this implies that in such an idealized context
(where the losses are zero) the TMS state is superior to the TF state for performing a quantum interferometry experiment.
Indeed, examination of Fig. 2 reveals that the phase neighbourhood around which the TMS exhibits a better behaviour is
very narrow. Overall, even in the absence of losses, the phase domain where sub-shot-noise interferometry can be observed is much
larger using TF states than using TMS states. We posit that this $\sqrt{2}$ factor advantage of the TMS state at $\phi_0=0$
is more likely a result of mathematical accident than a predictable outcome based on physical argument.

When losses are introduced, the optimal phase resolution continuously evolves towards a situation where TF states perform better,
regardless of the phase. This implies that, when focusing on the optimal phase resolution as a function of the number of particles,
one can find a crossing point between TF and TMS: numerically this crossing point exists for ${0.946 \lesssim \eta < 1}$
as it is shown in \cref{fig:delta_phi_phi0} below.

\begin{figure}
	\includegraphics[width=\linewidth]{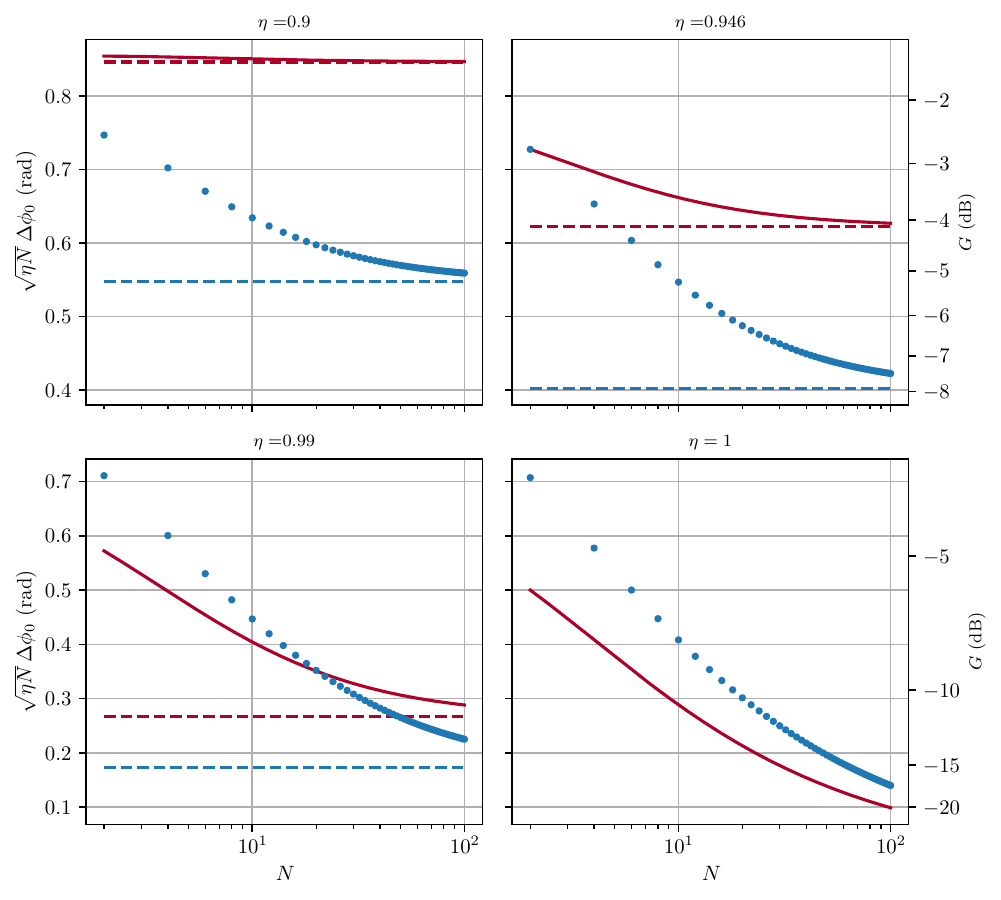}
	\caption{Asymptotic behaviour of the ratio between the phase uncertainty $\Delta \phi_0$ (i.e. $\Delta \phi$ estimated at
		the optimal phase $\phi_0$) and the SQL, as a function
		of the number of particles, for both TF (blue points) and TMS states (red lines), for various quantum efficiencies. 
		We also give the gain in decibel defined as $G = 20 \log \left(\sqrt{\eta N} \, \Delta\phi_0\right)$.
	}
	\label{fig:delta_phi_phi0}
\end{figure}